\documentclass[11pt]{article} 
\usepackage{epsfig}                   
\usepackage{color}

\begin{document}
\pagestyle{empty}
\begin{flushleft}
\vspace*{-1.8cm}
{\bf\hspace*{16cm}{  LAL 01-87}}\\
{\bf\small\hspace*{14cm}  PRHEP-HEP2001/261}\\
{\hspace*{15.5cm}{ December 2001}}\\
\end{flushleft}
\vspace*{4cm}

\begin{center}
{\LARGE\bf Electron polarization measurement\\ 
\vspace*{0.2cm}
using a Fabry-P\'erot cavity at HERA}
\end{center}
\vspace*{1cm}
\begin{center}
{\Large\bf Zhiqing Zhang}\footnote{
On behalf 
of E.~Barrelet, V.~Brisson,
M.~Jacquet-Lemire, 
A.~Reboux, C.~Pascaud, F.~Zomer\\  \hspace*{0.6cm} and the HERA POL2000
group.}
\end{center}

\begin{center}
{\large\bf Laboratoire de l'Acc\'el\'erateur Lin\'eaire} \\
IN2P3/CNRS et Universit\'e Paris-Sud - Bât. 200 - BP 34 - 91898 Orsay cedex, France	 
\end{center}

\vspace*{2cm}
\abstract{A new Compton longitudinal polarimeter currently under construction 
for HERA is presented. The key component of the polarimeter is a
Fabry-P\'erot cavity located around the electron beam pipe. With such an
optical cavity, a continuous laser power equivalent to 5\,kW, much higher than 
those commercially available, can be achieved, leading to one backscattered 
photon per bunch crossing. This ``few-photon mode'' will
allow a very precise determination of the calorimeter response with little
systematic uncertainty. The electron polarization measurement at the per
mill level is expected.}

\vspace*{3cm}
\begin{center}
{\it Talk given at the International Europhysics Conference on High Energy Physics\\
Budapest (Hungary), July 12-18, 2001}
\end{center}

\newpage
\pagestyle{plain}

\section{Introduction}
The study of deep inelastic scattering of polarized electrons with protons
at very high energies will be one of the major new physics topics after 
the upgrade of the HERA storage ring. The colliding-beam experiments 
H1 and ZEUS will for the first time have access to longitudinal polarized 
electrons\footnote{In this paper, ``electron'' is generically used to 
refer to both electron and positron.}.
The longitudinal polarization of the electron beam in a storage ring like HERA 
is achieved with the spin rotators from the naturally polarized transverse 
polarization through the emission of synchrotron radiation which is known as 
the Sokolov-Ternov effect~\cite{st}.

Currently the transverse and longitudinal electron polarizations at HERA 
are measured with, respectively, a transverse polarimeter (TPOL) and 
a longitudinal polarimeter (LPOL). The new LPOL will use a Fabry-P\'erot 
cavity to stock photons from a laser in resonance such that an amplified
laser beam can be obtained. Such a technique has been successfully employed 
at CEBAF~\cite{cebaf}.

The paper is organized as follows. In section \ref{sec:necessity}, the
necessities for having a fast and precise polarization measurement are briefly
discussed. The new LPOL is presented in section \ref{sec:newlpol}.

\section{The necessities for having a fast and precise polarization
measurement}\label{sec:necessity}
The basic HERA physics program including using polarized electron beams
has been described in a series of workshops~\cite{heraws}.
Here we shall only briefly discuss why it is important to have a fast and
precise electron polarization ($P_e$) measurement.

\begin{figure}[h]
\begin{center}
\epsfig{figure=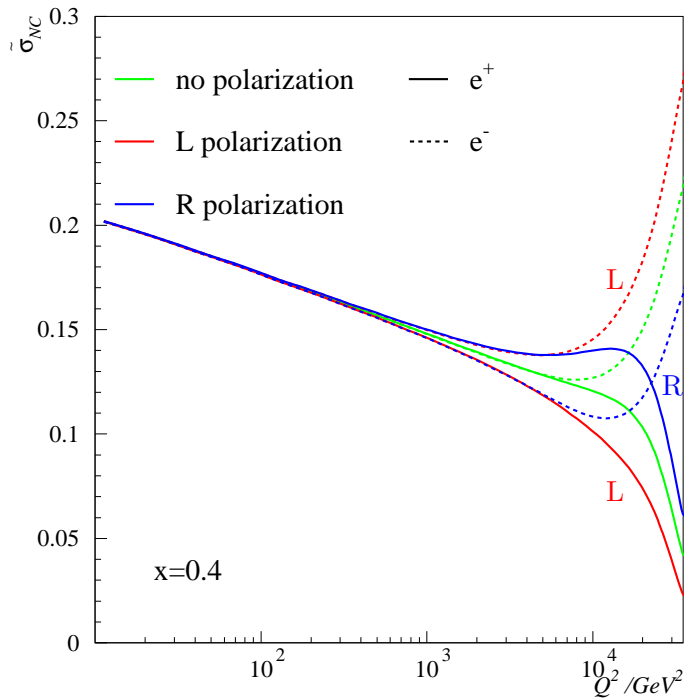,width=9 cm}%
\begin{picture}(50,100)
\put(-30,155){\textcolor{red}{L}}
\put(-9,115){\textcolor{blue}{R}}
\put(-30,75){\textcolor{red}{L}}
\end{picture}
\begin{minipage}[b]{17cm}
\caption{\label{fig:nc}The reduced $e^\pm p$ neutral current (NC)
cross-section as a function of $Q^2$ for different electron polarizations.
The cross-section is largest (smallest) for left(L) polarization
in the $e^-p$ ($e^+p$) collisions.}
\end{minipage}
\end{center}
\end{figure}

As far as the colliding experiments H1 and ZEUS are concerned, the dominant
Standard Model processes, neutral current (NC) and charged current (CC)
interactions, are very sensitive to the electron beam polarization at high
$Q^2$ as it is shown for the NC case in figure \ref{fig:nc}.%
The NC cross-sections can be expressed in terms of structure functions
$\tilde{F}_2$ and $x\tilde{F}_3$, which contain linear combinations of the
quark densities and quadratic functions of the vectorial ($v_q$) and axial
($a_q$) couplings of the quarks ($q$) to the $Z$ boson.
A precise knowledge of $P_e$ is thus necessary in order to measure accurately 
the NC cross-sections, which in turn provide a unique constraint on the
parton densities at large $x$~\cite{ringberg01}. The same cross-section data
can also be used to measure the coupling constants $v_{u,d}$ and $a_{u,d}$ with
a precision comparable to those obtained by LEP for the couplings of the
heavy quarks. The polarized beam is necessary for determining the vector
couplings $v_{u,d}$ and desirable for enhancing the sensitivity for $a_{u,d}$.

Similarly, the reduced CC cross-sections $\tilde{\sigma}_{\rm CC}(e^+p)=
(1+P_e)x\left[(\overline{u}+\overline{c})+(1-y)^2(d+s)\right]$ and
$\tilde{\sigma}_{\rm CC}(e^-p)=
(1-P_e)x\left[(u+c)+(1-y)^2(\overline{d}+\overline{s})\right]$
can be larger or smaller than the unpolarized cross-sections.
A larger cross-section will effectively improve the statistical precision in
probing the helicity structure of the CC interaction and in determining the
propagator mass of the exchanged $W$ boson.
The CC cross-sections data also provide a means to search directly 
for right-handed CC interactions. The search sensitivity depends, however,
crucially on the precision of the electron polarization measurement.

Although the current polarimeters do not allow a precise and fast
polarization measurement on a bunch-by-bunch base, there is indication that
there are significant polarization variations both within a luminosity fill
as a function of time and 
among different bunches (e.g.\ between colliding and non-colliding bunches).
A fast measurement will thus provide the HERA machine a quick feedback to
optimize the polarization and maintain the maximum polarization.

\section{The new longitudinal polarimeter}\label{sec:newlpol}
The principle of the polarization measurement relies on the non-destructive
method of detecting backscattered photons produced in Compton scattering of an
intense circularly polarized laser (photon) beam off the electron beam. The
transverse polarization is measured from the vertical asymmetries between
the left and right circularly polarized photons, whereas the
longitudinal polarization is determined from the energy symmetries of the
different photon helicities. The
current statistical precision of the LPOL is around 1\% for 2 minutes of
data taking averaged over all bunches in the ring (up to 190). The
systematic uncertainty is about 2\% which is dominated by the energy scale
uncertainty of the calorimeter measuring the backscattered photons in
multi-photon mode\footnote{\hspace*{0.5cm}\parbox{17cm}{Three different modes of operation can be
distinguished for a polarimeter detecting the backscattered photons:
(1) single-photon mode: The probability to scatter a photon is very low
(0.01 for the existing TPOL), so that at most one phone per interaction 
is detected in the calorimeter), (2) few-photon mode: The probability 
is high enough that on average one photon per bunch crossing is observed,
(3) multi-photon mode: Per bunch crossing a large number of photons 
is scattered into the calorimeter (e.g.\ around 1000 for the actual LPOL).}}. 
To reach a same statistical precision for a single bunch, 
more than 30 minutes are required due to the low pulsed laser frequency 
(up to 100\,Hz to be compared with the electron bunch
repetition rate of 10\,MHz). The statistical precision of the TPOL is 1\% 
per minute again averaged over all bunches (the current TPOL data acquisition 
(DAQ) system does not allow the differentiation of bunch to be made). 
The systematic uncertainty is about twice that of the LPOL due mainly to 
the limited statistics of the dedicated rise-time runs taken with 
a flat machine and needed to define the absolute polarization scale.

The new LPOL will be installed next to the existing LPOL
during the shutdown early next year. The new LPOL is expected to provide the
most precise measurement of the polarization with a statistical error of
0.1\% (0.3\%) for the colliding (non-colliding) bunches and 1\% per minute
for every single bunch, and a systematic uncertainty at per mill level.

\begin{figure}
\begin{center}
\epsfig{file=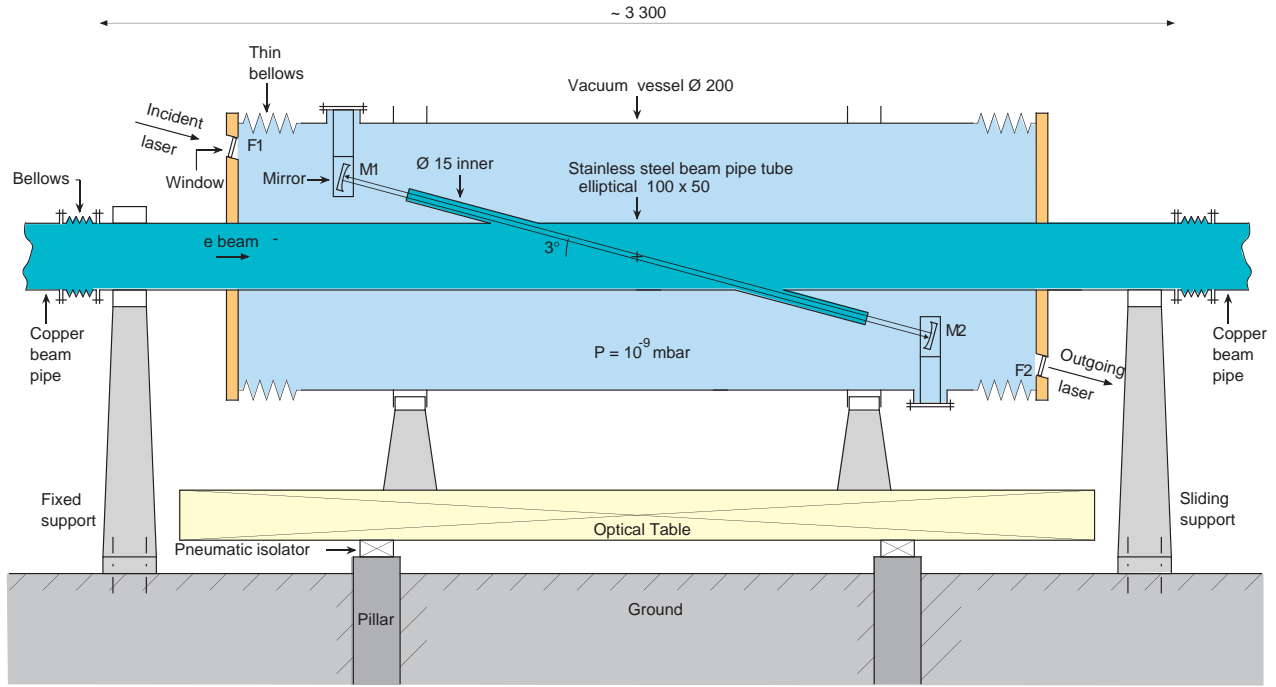,width=17cm}%
        \caption{Schematic layout of the new longitudinal polarimeter.}%
	\label{fig:cavity}
\end{center}
\end{figure}

The conceptual layout of the new LPOL is shown schematically in 
figure \ref{fig:cavity}.%
The cavity surrounds the interaction point for a length of 2\,m. The main
part of the cavity is a cylindrical vacuum vessel made of stainless steel, 
with a wall
thickness of around 0.3\,cm. The optical elements (mirrors) of the cavity are
mechanically decoupled from the beam pipe and are rigidly
fixed to a stable optical table. The latter is isolated from the HERA tunnel
through pneumatic isolators. The central part of the cavity is connected to
two end pieces by means of two bellows. Only the end pieces are in contact
with the beam pipe. The bellows, acting as a vibration filter, block thus
vibrations in the machine from reaching the mirrors. 

In addition to the Fabry-P\'erot cavity, there are a few other optical
components mounted on the optical table, in particular the infrared laser
source of 0.7\,W (Nd:YAG, $\lambda=1064$\,nm). The role of the these
external optical elements is twofold: the laser beam after having been
highly circularly polarized has to be transported into the cavity, 
and the properties of the laser beam have to be measured and monitored.

The whole system is placed in a hut maintaining a stable temperature
within $1^\circ$C so that the resulting mechanical deformation of
the cavity can be handled by an electronic feedback system acting on the
frequency of the tunable laser.

To estimate the precision with which the polarization can be measured with
the new device, a number of systematic effects have been studied: 
1) the energy scale and energy linearity of the calorimeter, 
2) the fluctuations of the laser power inside the cavity, 
3) the dead material in front of the polarimeter,
4) the properties of the electron beam, and
5) the polarization precision of the laser light.
Based on the simulations and the experience obtained from the similar cavity 
at CEBAF, all considered effects can be
controlled such that the systematic uncertainty of the electron
polarization would be measured at per mill level. In particular, a precise
in-situ energy calibration is possible under the few-photon operation mode,
in contrast with the multi-photon mode, since the Compton spectra of one or 
two photons are still visible in the measured 
energy distribution. 

\section{Summary}
The new longitudinal polarimeter being constructed will use a Fabry-P\'eroy
cavity to significantly gain the effective power of a laser light. The
increased power will enable us to operate the polarimeter in the few-photon
mode, and still reaching a high statistical precision of ${\cal O}(0.1)$\%.
The new polarimeter has a number of important advantages over the existing
ones:
\begin{itemize}
\item The polarimeter can be calibrated using the same data as for the
polarization measurement.
\item Systematic uncertainties can be tracked and controlled based on the
information obtained from the data.
\item Fast and reliable feedback can be given to the machine, helping to
optimize the polarization.
\end{itemize}

The precision that the device is expected to reach will be around
$0.2-0.4$\% on the polarization. This will guarantee that the physics
potential of HERA is not limited by the polarimetry, and that high precision
measurements for the study of neutral and charged currents are possible.

\end{document}